\documentclass[twocolumn,epjc3]{svjour3}

\RequirePackage{graphicx}
\RequirePackage{mathptmx}      
\RequirePackage{flushend}
\RequirePackage{color}
\usepackage{amsmath}
\usepackage{hyperref}
\usepackage{color}
\usepackage[dvipsnames]{xcolor}
\usepackage[section]{placeins}
\usepackage[normalem]{ulem}
\usepackage{appendix}

\journalname{Eur. Phys. J. A}

\begin{document}

\title{A global study of $\alpha$-clusters decay in heavy and superheavy nuclei with half-life and preformation factor}

\author{G. Saxena$^{1,2}$, P. K. Sharma$^{3}$, Prafulla Saxena$^{4}$}

\institute{
\mbox{} \\[-8bp]
\mbox{$^{1}$Department of Physics (H $\&$ S), Govt. Women Engineering College, Ajmer-305002, India}\\
\mbox{$^{2}$Department of Physics, Faculty of Science, University of Zagreb, HR-10000 Zagreb, Croatia}\\
\mbox{$^{3}$Govt. Polytechnic College, Rajsamand-313324, India}\\
\mbox{$^{4}$Department of Computer Science and Engineering, Malviya National Institute of Technology, Jaipur-302016, India}\\
\\
\mbox{e-mail: gauravphy@gmail.com}\\
}

\authorrunning{G. Saxena \textit{et al.}....}
\titlerunning{A global study of $\alpha$-clusters decay in heavy and superheavy nuclei....}

\date{\today}


\maketitle

\begin{abstract}
A detailed study of $\alpha$-clusters decay is exhibited by incorporating crucial microscopic nuclear structure information into the estimations of half-life and preformation factor. For the first time, using the k-cross validation approach, two semi-empirical formulas for (i) $\alpha$-decay half-life and (ii) $\alpha$-particle preformation factor, are picked out and subsequently modified by including shell, odd-nucleon blocking, and asymmetry effects along with the usual dependence on $\alpha$-decay energy ($Q_{\alpha}$) and angular momentum of $\alpha$-particle. Both the formulas are fitted for the two different regions separated by neutron number N$=$126, as from the experimental systematics the role of N$=$126 shell closure is found decisive in determining the trends of $Q_{\alpha}$, $\alpha$-decay half-life, and $\alpha$-particle preformation factor. It is found that the inclusion of the above-mentioned degrees of freedom significantly reduces the errors in the estimations when compared with several other similar modified/refitted semi-empirical relations indicating the robustness of the proposed formulas. The predictions of $\alpha$-decay half-life throughout the periodic chart have been made including the unknown territory, future probable decay chain of self-conjugate nucleus $^{112}$Ba terminated on $^{100}$Sn, decay chain of $^{208}$Pa through new isotope $^{204}$Ac as well as decay chains of awaiting superheavy nuclei $^{298}$Og and $^{299}$120.
This article is expected to provide a systematic approach to selecting the formula by which reliable predictions can be made.

\end{abstract}

\section{Introduction}
Starting from the identification of $\alpha$-particle radioactivity in 1907 and its explanation based on (a) tunneling in 1928 by George Gamow \cite{gamow1928}, and
(b) laws of quantum mechanics by Gurney \& Condon \cite{gurney1928a}, for almost a century this decay mode has been serving as a coherent pathway for the development of theoretical and experimental treatments in nuclear physics research. Recent observations of (i) superallowed $\alpha$-decay to doubly magic $^{100}$Sn \cite{auranen2018}, (ii) direct experimental evidence for the formation of $\alpha$-clusters at the surface of neutron-rich tin isotopes \cite{tanaka2021}, (iii) $\alpha$-decay in a new proton-unbound isotope $^{204}$Ac \cite{huang2022}, (iv) $\alpha$-decay chains of $^{293}$117 and $^{294}$117 with 11 new nuclei \cite{og2010} (v) $\alpha$-decay chains of the heaviest element $^{294}$118 \cite{ogan2006}, (vi) eleven new heaviest isotopes of elements Z$=$105 to Z$=$117 \cite{ogan2011}, as well as attempts of synthesizing new elements with Z$=$119 and Z$=$120 \cite{voinov2020} are few of the significant outcomes of advancements of various experimental facilities at Argonne National Laboratory in USA \cite{davids1989}, GSI in Germany \cite{hofmann2000,hofmann2011}, RIKEN in Japan \cite{morita2007}, and Dubna laboratory in Russia \cite{og2010,og2015npa}, etc. based on the detection of $\alpha$-particle emission. The more progressive analysis of these experimental facilities and their results can be found in the recent reviews \cite{nazar2018,giuliani2019}.\par

To alleviate any experimental plan the theoretical inputs become oftentimes pivotal, and for the $\alpha$-decay study these inputs can be broadly categorized into two parts: (i) estimation of probable reactions and their cross-sections \cite{og2015npa,giardina2018} (ii) prediction of the half-lives and the decay modes for various unknown nuclei. The latter category can be further subdivided into two methods out of which the first one is based on the computation of $\alpha$-decay half-lives by employing various theoretical methods or models such as the Gamow-like model (GLM) \cite{Zdeb2013}, fission-like model \cite{yong2010}, liquid drop model \cite{poenaru1979} with their modifications \cite{cui2018,bao2014,royer2008}, Coulomb and proximity potential model (CPPM) \cite{zanganah2020NPA,santhosh2012NPA}, etc. The second powerful tool to calculate $\alpha$-transition probabilities or decay width of cluster channel relies on the nuclear shell model \cite{mang1960,sandulescu1964,soloviev1962,mang1964} or microscopic formalism \cite{delion1992,delion1994,delion2000}. With the use of the shell model the experimental data corresponding to a large number of $\alpha$-decay transitions are well reproduced \cite{fliessbach1976,tonozuka1970,lenzi1993} including the computation of the preformation factor and penetrability of $\alpha$-emission with complex-energy shell model \cite{betan2012}. On the other hand, the microscopic formalism is used to estimate the $\alpha$–particle formation probability, and, recently, by one of the microscopic formalisms i.e Hartree-Fock-Bogoliubov (HFB) method the experimental values of $\alpha$-decay widths are described accurately \cite{dumitrescu2023}.\par

An alternate for finding the $\alpha$-decay half-lives belongs to the use of (semi)empirical formulas \cite{vss1966,sobi1989,parkho2005,brown1992,renA2004,qi2009,royer2000,saxena2021} or their modified(refitted) versions \cite{xu2022,akrawy2017,akrawy2019,singh2020,Saxena2021jpg,akrawy2022EPJA,sharma2021npa} mainly based on Geiger-Nuttall law. So far more than 50 empirical/semi-empirical/ modified versions of the formulas are available in the literature to estimate $\alpha$-decay half-life. Looking into these several formulas and at the pitfalls and common misconceptions of empirical formulas \cite{beckerley1945}, the aim of the present article is not to add another one to the list. In fact, the objective of this report is to put various versions of the formulas on equal footing with the same data-set (by refitting) to prove their credibility on the grounds of k-cross validation, and if needed then to improve the estimation of half-lives by incorporating imperative microscopic nuclear structure information.\par

Another aim of the present article is to bring the $\alpha$-particle preformation factor, which corresponds to the probability of the formation of an $\alpha$-cluster within the surface of a parent nucleus, into the mainstream together with the half-lives estimation. In the past, it has been shown that due to the complicated structure of quantum many-body systems as heavy (82$\leq$Z$\leq$102) and superheavy nuclei (Z$\geq$103), the determination of the preformation factor is quite difficult \cite{delion1996,lovas1998}. Also, there are a few empirical formulas available \cite{zhang2008,zhang2009,deng2017,deng2018,liu2020cpc,wang2020,deng2022cpc} to calculate preformation factors based on valence nucleons (holes) corresponding to the known shell closures, however, their applicability in the mid-shell and superheavy region is rather questionable. There are, however, only a few formulas \cite{delion2009,ni2010,deng2020,santhosh2021} which consider $Q$-value dependency and result reasonably well in the mentioned regions. But, at the same time, these formulas neglect the important shell and odd-even effects. Recently, Deng and Zhang \cite{dengplb2021} have proposed another modified version of their earlier formula \cite{deng2020} by considering the missing effects in addition to the $Q$-value dependency. However, there is still an obvious requirement to check the other microscopic dependencies on the $\alpha$-particle preformation factor similar to the empirical relationships of the $\alpha$-decay half-life, which is precisely the objective of the present work.\par

Hence, this investigation is three-fold leading to (i) examine of the credibility of available formulas utilizing k-cross validation, (ii) propose the formulas of $\alpha$-decay half-life as well as preformation factor considering crucial microscopic nuclear structure information, and, (iii) evaluate the half-lives and preformation factor for the unknown nuclei along with future probable decay chains.


\section{Cornerstone for formulas}
The closed shell nuclei with magic numbers $Z=50,82,$ and $N=82,126$ are the ones that substantially affect the decay properties of nuclei towards the heavier region of the periodic chart. In other words, the characteristics of such nuclei provide a key basis on which several models or empirical relations linked to the decay mechanism are constituted. As an example, the decay of $^{212}$Po into $^{208}$Pb led to Gamow's theory in which the $\alpha$-decay process was explained by the penetration of a preformed $\alpha$-particle through the Coulomb barrier which is usually formulated as linear dependence on the charge number of the daughter nucleus and reciprocal of the square root of $\alpha$-decay energy ($\sqrt{Q}$) for $\alpha$-decay half-life \cite{geiger1911,vss1966}. Various other dependencies on $\alpha$-decay half-life were articulated from time to time in terms of few additional terms or fitting with the available data sets \cite{vss1966,sobi1989,parkho2005,brown1992,renA2004,qi2009,royer2000,saxena2021,xu2022,akrawy2017,akrawy2019,singh2020,Saxena2021jpg,akrawy2022EPJA}.
In a similar way, the probability of an $\alpha$-cluster formation inside its parent nucleus is hypothesized in terms of the $\alpha$-particle preformation factor which is empirically expressed by various formulas based on valence nucleons (holes) corresponding to the known shell closures or alternatively only by a few formulas \cite{delion2009,ni2010,deng2020,santhosh2021,dengplb2021} based on $Q$-values or shell effects, or both.\par
\subsection{Formula for $\alpha$-decay half-life}
It will be worthwhile to check the validity of already established and widely used empirical formulas of $\alpha$-decay half-life for the considered latest data-set of 349 nuclei \cite{audi2020}, before constituting any other empirical formula. One common and easiest way to compare all the formulas is to compute the $\alpha$-decay half-life using the known $Q_{\alpha}$-values \cite{audi2020} along with the provided values of coefficients of respective formulas from their original literature. However, this method carries a major shortcoming that the provided values of coefficients were obtained by different data sets available at the time of the introduction of the respective formula. This flaw, still, can be rooted out by refitting all the concerned formulas on the same dataset (349 in the present paper) and determining the new coefficients. The second way allows bringing all the formulas on equal footing, however, it still comprises a problem of training (fitting) and testing on the same dataset which may give rise to overfitting and in principle may not be reliable for the unknown data(nuclei). The best way to decide the reliability of a given formula on unseen (unknown) data is to separate out training data and the test data, which can eventually incur with the use of the k-cross validation method: a very effective method primarily used in applied machine learning to estimate the skill of a machine learning model on unseen data, and now recently applied in a few nuclear physics studies \cite{carnini2020,li2022}.

\begin{table*}[!htbp]
\caption{Root mean square errors (RMSE) for various formulas using different sets of data for fitting and testing (see the text for details). The second last column contains the average value of RMSE for all 5 sets. The last column lists the value of $\chi^2$ per degree of freedom computed by using the total number of fitted coefficients for the respective formula.}
\centering
\resizebox{1.0\textwidth}{!}{%
\begin{tabular}{l@{\hskip 0.5in}|c@{\hskip 0.5in}c@{\hskip 0.5in}c@{\hskip 0.5in}c@{\hskip 0.5in}c|@{\hskip 0.5in}c@{\hskip 0.5in}|c}
\hline
\hline
\multicolumn{1}{l}{Formula for}&\multicolumn{6}{|c|}{RMSE}&\multicolumn{1}{c}{$\chi^2$}\\
\cline{2-7}
$\alpha$-decay half-life       &  set 1 & set 2 & set 3 & set 4 & set 5& Average & \\
\hline
NMTN 2023 (present work)                        &  0.3999 &0.4614	&0.3856	&0.4244	&0.4150	&0.4172 &   0.15  \\
Soylu 2021 \cite{soylu2021}                   &  0.4401	&0.4595	&0.3749	&0.4037	&0.4162	&0.4189     &  0.61     \\
Royer 2020 \cite{royer2020}                   &  0.4911	&0.4593	&0.3895	&0.4166	&0.4124	&0.4338     &  0.24     \\
MRenB 2019 \cite{newrenA-mrenB2019}            &  0.5009	&0.4899	&0.3965	&0.4287	&0.4290	&0.4490 &  0.67    \\
DK2 2018  \cite{DK2}                          &  0.4154	&0.6266	&0.3971	&0.4132	&0.4293	&0.4563     &  0.50        \\
NMHF 2021 \cite{sharma2021npa}                 &  0.5046	&0.5462	&0.3983	&0.4134	&0.4323	&0.4590 &  0.51    \\
MTNF 2021 \cite{saxena2021}                   &  0.4915	&0.4795	&0.4288	&0.4841	&0.4411	&0.4650     &  0.49        \\
MRF 2018  \cite{MRF}                           &  0.5006	&0.5062	&0.4350	&0.4183	&0.4729	&0.4666  & 1.05    \\
AAF 2018 \cite{akrawyijmpe2018}                &  0.4725	&0.5313	&0.4582	&0.4446	&0.5040	&0.4821  & 3.20    \\
MVS 2019 \cite{akrawyprc2019}                 &  0.6345	&0.6647	&0.5272	&0.5594	&0.5382	&0.5848      & 0.47       \\
NRenA 2019 \cite{newrenA-mrenB2019}           &  0.6527	&0.6404	&0.5284	&0.5625	&0.5441	&0.5856      & 0.40       \\
MUDL 2019 \cite{akrawy2019}                   &  0.6618	&0.6406	&0.5281	&0.5631	&0.5478	&0.5883      & 0.45       \\
QF 2021 \cite{saxena2021}                     &  0.6280	&0.7204	&0.5633	&0.6360	&0.5552	&0.6206      & 0.36       \\
MSLB 2019 \cite{akrawyprc2019}                &  0.6082	&0.8145	&0.5757	&0.6321	&0.5460	&0.6353      & 0.54       \\
NMMF 2021 \cite{sharma2021npa}                 &  0.8238	&0.8407	&0.8357	&0.9350	&0.7306	&0.8331  & 0.87    \\
IRF 2022 \cite{ismail2022}                    &  2.3761	&2.3506	&1.4389	&1.3929	&1.1667	&1.7450      & 0.37       \\
\hline
\hline
\end{tabular}}
\label{alpha-rmse}
\end{table*}

Such a method can be used here to estimate how the model (formula) is expected to perform in general when used to make predictions on data not used during the training of the model (formula). In addition, the method ensures that each data sample gets a chance to appear in the train set as well as the test set. This method is very popular and simple to understand which generally results in a less biased or less optimistic estimate of the model skill than other methods. \par

With the k-cross approach, the robustness of the employed method is ensured by considering all data points as test samples in different splits rather than a specific test split which might luckily give better performance. The method has a single parameter called k that refers to the number of groups that a given data sample is to be split into by which the overfitting chances can be avoided. Typically, the k value in a dataset depends on several factors, such as the bias-variance trade-off, computational cost, data characteristics, etc. There's no single best-fit answer for the k value as it can vary from k = 2 to n (total number of data), but it often depends on a trade-off between accuracy, variance, and computational efficiency. A smaller value of k, like 2 or 3, can lead to higher variance in test error as the training set becomes smaller and more variable, whereas a larger k value can increase computational costs and result in fewer test samples in smaller datasets. The common practice is to choose a k value between 5 and 10 \cite{k-cross2010} which we have also experimented with and led to the optimal solution at k=5. Hence, in the present case, we use a 5-fold cross-validation for the 349 data-set used. In this way, we first divide the full data-set of 349 data into set 1 of 280 data (for fitting) and test 1 of 69 data (for testing) and so on for other sets (set 2, set 3 ...etc.). It is ensured that the data in fitting sets are different than the data in testing sets and contain the distributed data in the entire range i.e. 52$\leq$Z$\leq$118. We have considered several recently reported formulas and first fitted them on 5 different sets and then probed on the completely unseen 5 different respective test data-sets. The root mean square errors (RMSE) as mentioned in Eqn. (1) (where $N_{d}$ is the number of data points) in estimating $\alpha$-decay half-life for the test data-sets are mentioned in Table \ref{alpha-rmse} in front of the respective formulas along with the average values (for all 5 sets) of RMSE.
 \begin{eqnarray}
    RMSE &=& \sqrt{\frac{1}{N_{d}}\sum^{N_{d}}_{i=1} (logT^i_{Th}-logT^i_{Expt.})^2}
 \end{eqnarray}
  \begin{eqnarray}
    \chi^2 &=& \frac{1}{N_d-N_p} \sum_{i=1}^{N_d} (logT^i_{Th}-logT^i_{Expt.})^2
\end{eqnarray}
The Table consists of the modified versions of the universal decay law represented by Soylu 2021 and MUDL 2019 from Refs. \cite{soylu2021} and \cite{akrawy2019}, respectively, the Royer formula represented by Royer 2020 \cite{royer2020} along with its modified and improved versions represented by AAF 2018 \cite{akrawyijmpe2018},  MRF 2018 \cite{MRF} and IRF 2022 \cite{ismail2022}, new Ren A formula (NRenA 2019) \cite{newrenA-mrenB2019}, modified Ren B formula (MRenB 2019) \cite{newrenA-mrenB2019}, new modified Horoi formula (NMHF 2021) \cite{sharma2021npa}, new modified Manjunahta formula (NMMF 2021) \cite{sharma2021npa}, quadratic fitting formula (QF 2021) \cite{saxena2021}, modified Tagepera and Nurmia formula (MTNF 2021) \cite{saxena2021}, modified Denisov and Khudenko formula (DK2 2018) \cite{DK2},  the modified version of scaling law of Brown (MSLB 2019) \cite{akrawyprc2019}, and modified version of Viola and Seaborg formula (MVS 2019) \cite{akrawyprc2019} for comparison.
The list can further be made longer by including many other formulas but we have considered the recent ones as well as those in which dependency on angular momentum ($l$) of the emitted $\alpha$-particle is already incorporated for an accurate consideration of the $\alpha$-transitions \cite{bohr1975} and to describe favoured and unfavoured transitions, distinctly  \cite{denisov2009}.\par

\begin{figure}[!htbp]
\centering
\includegraphics[width=0.5\textwidth]{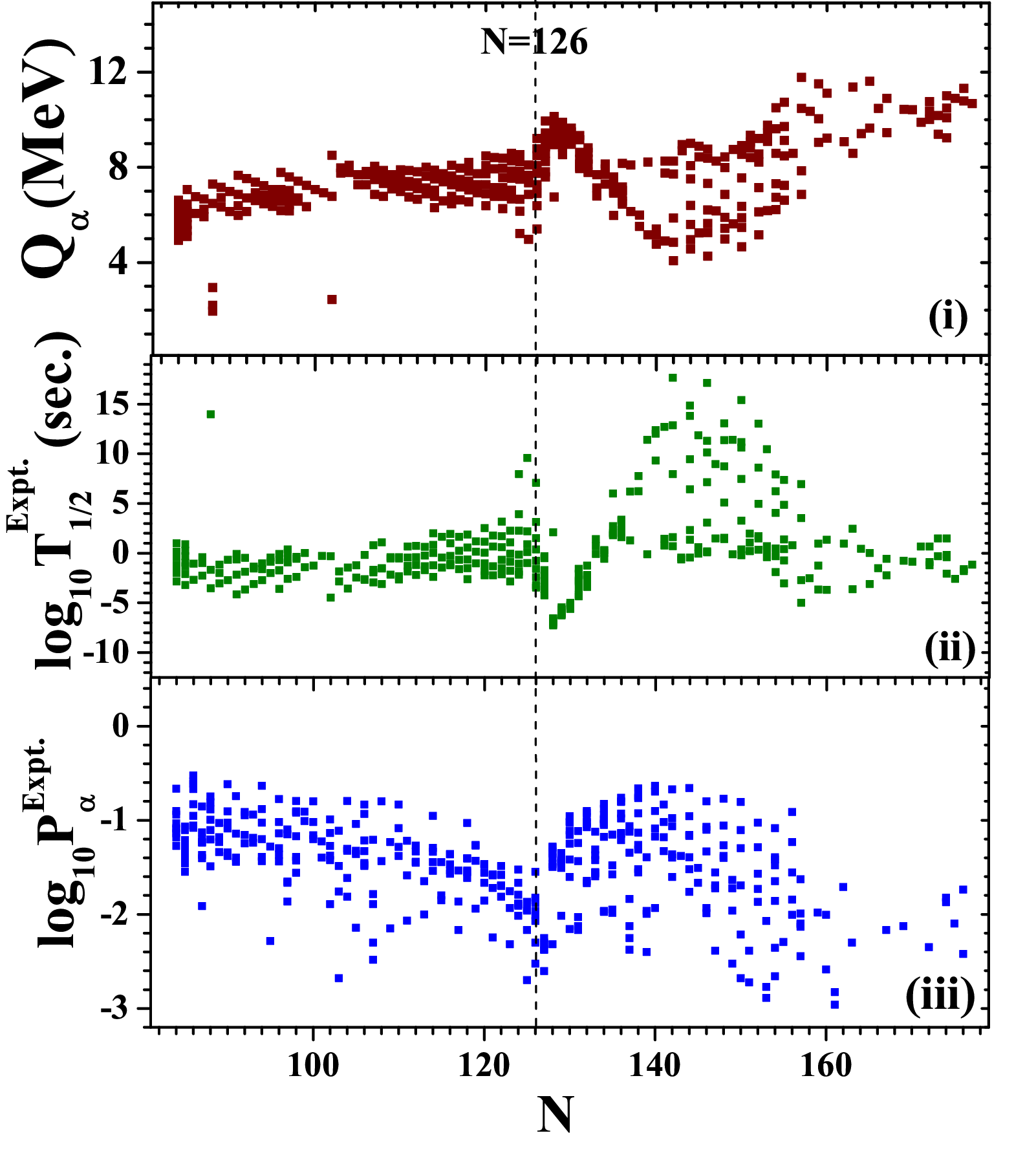}
\caption{(i) $\alpha$-decay energy ($Q_{\alpha}$), (ii) $\alpha$-decay half-life ($log_{10}T_{1/2}^{Expt.}$), and (iii) $\alpha$-particle preformation factor ($log_{10}P_{\alpha}^{Expt.}$) of 349 nuclei \cite{audi2020}.}\label{Fig:126}
\end{figure}

It is clear from Table \ref{alpha-rmse} that the formula given by Soylu \textit{et al.} \cite{soylu2021}, comprising isospin ($I=(N-Z)/A$) and angular momentum ($l$) dependencies, results with minimum RMSE while compared to all the formulas for the same class which in the course of time indicates the prerequisite of these two structural properties. Now, to observe the effect of another structural property i.e. closed shell at $N=126$, on the $\alpha$-decay half-life, we plot known $\alpha$-decay energy ($Q_{\alpha}$) as well as known $\alpha$-decay half-life ($log_{10}T_{1/2}^{Expt.}$) of 349 nuclei taken from NUBASE2020 \cite{audi2020} as a function of the neutron number of parent nuclei in Figs. \ref{Fig:126} (i) and (ii), respectively. A very sharp discontinuity in the systematic trend of Q-values and $log_{10}T_{1/2}$ values around $N=126$ can be observed in the figures. This sharp change separates the set of nuclei into two regions (i) $N\leq126$ (ii) $N>126$ and indicates the need for a formula of $\alpha$-decay half-life incorporating the two regions, separately.\par

\begin{figure}[!htbp]
\centering
\includegraphics[width=10.2cm, height=7.6cm]{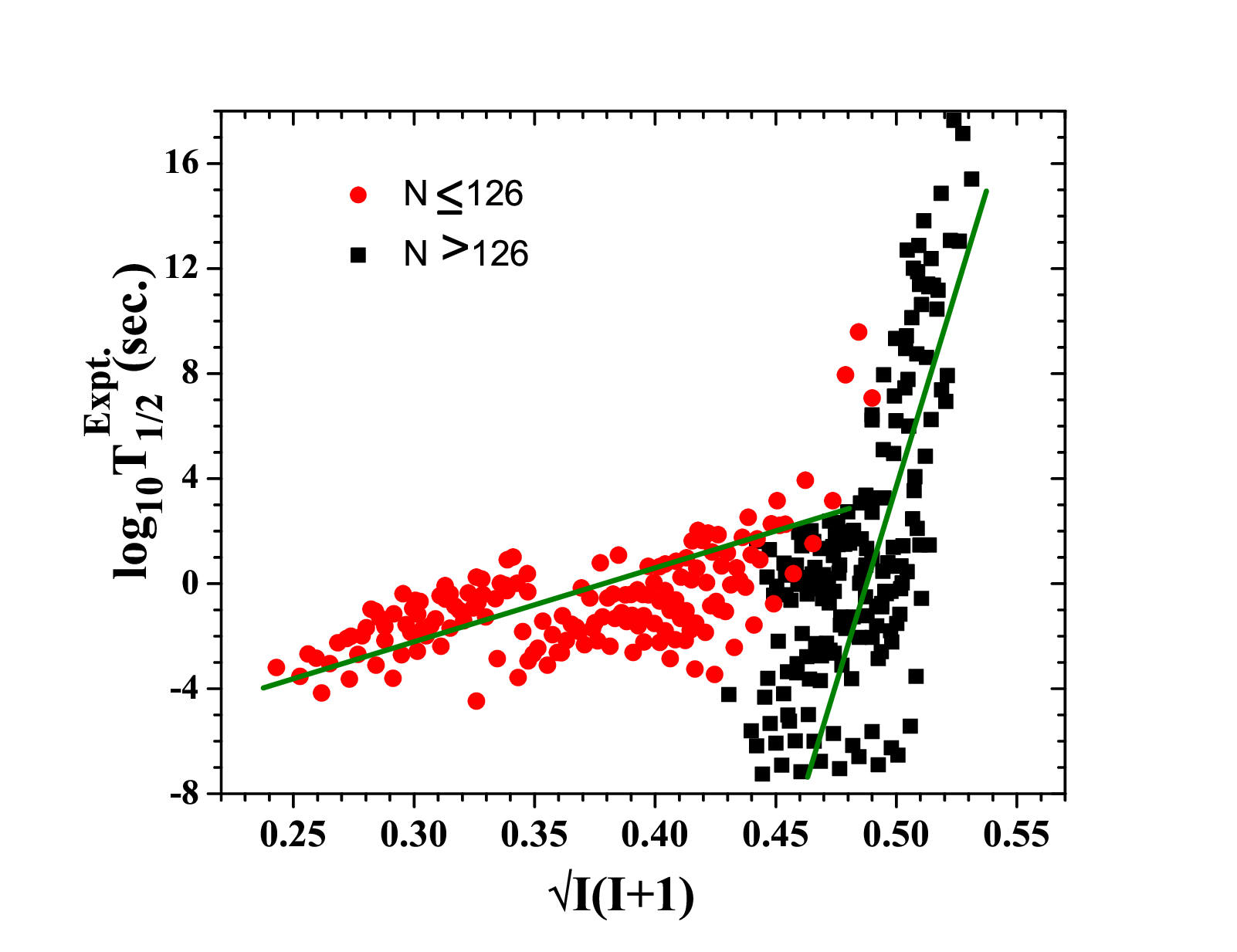}
\caption{$\alpha$-decay half-life ($log_{10}T_{1/2}^{Expt.}$) of 349 nuclei \cite{audi2020} with $N\leq126$ (red circle) and $N>126$ (black square) as a function of $\sqrt{I(I+1)}$; where $I=(N-Z)/A$.}
\label{Fig:I}
\end{figure}

In addition to the imprint of a closed shell, another factor that may affect the decay properties of nuclei is the isospin asymmetry of the parent nucleus ($I=(N-Z)/A$). As per the Soylu \textit{et al.} \cite{soylu2021} formula, $\alpha$-decay half-lives are found sensitive to this factor. Therefore, the dependence of isospin asymmetry is systematically investigated with 349 nuclei in Fig. \ref{Fig:I} for the two regions (i) $N\leq126$ (ii) $N>126$ in which known $\alpha$-decay half-lives \cite{audi2020} are plotted as a function of $\sqrt{I(I+1)}$; where $I=(N-Z)/A$. The variation of $\alpha$-decay half-lives with $\sqrt{I(I+1)}$ evidently demonstrates its significance. In addition, the divergent variation for the nuclei $N\leq126$ and $N>126$ suggests a separate fitting for both regions. It is important to point out here that the region for the value of $\sqrt{I(I+1)}$ $\sim$ 0.35 is related to the nuclei in which N is close to 126 and that corresponds to region III of odd clustering behavior reported by Delion \textit{et al.} \cite{delion2020}. \par

With the above investigation, to include all the specified indispensable microscopic properties, we set up a new formula for $\alpha$-decay half-life. For the present case, the usual dependency on $Z_{d}/\sqrt{Q}$ can be chosen empirically which we opt from the Tagepera and Nurmia formula (TN formula) given in 1961. The old TN formula $log T_{1/2}(s) = a(Z_{d}Q^{-1/2}-Z_{d}^{2/3})+b+h_{oe}$ expresses $\alpha$-decay half-life in terms of the atomic numbers of the daughter nucleus and the Q-value of the two-body disintegrating system. This formula can be modified by considering the angular momentum of the emitted $\alpha$-particle which requires spins and parities of the parent and daughter nuclei. This consideration is found to delineate all set of experimental data, characterizes favoured and unfavoured transitions based on the standard selection rules \cite{denisov2009}, and already acquired as a centrifugal term in several proposed or modified versions of empirical formulas \cite{soylu2021,akrawy2019,royer2020,akrawyijmpe2018,MRF,ismail2022,newrenA-mrenB2019,sharma2021npa,saxena2021,DK2,akrawyprc2019,saxenaepja2023}. Another crucial consideration is the effect of unpaired nucleons which was eventually found to affect the half-lives while investigated separately in even-even, odd-odd, and odd-A nuclei. This is a signal of the existence of a pairing correlation in the nucleus as nucleons favour to pair and, therefore, $\alpha$-emission in even-even nuclei is more probable compared to odd-A and odd-odd nuclei. This effect of unpaired nucleons is referred to as the odd-even staggering effect \cite{royer2020} similar to the odd-even staggering effect in the binding energy of nuclei taken into account in Refs. \cite{schuck1973,gambhir1983}.\par

Summarily, we use the TN formula for $Z_{d}$ and $Q$ dependence and additional terms linked with microscopic nuclear structure information like (i) angular momentum ($l$) of the emitted $\alpha$-particle, (ii) isospin asymmetry ($I$) of parent nucleus, and (iii) the effect of unpaired nucleons, to propose the following semi-empirical formula for $\alpha$-decay half-life:
\begin{eqnarray}\label{NTN}
 log_{10}T_{1/2}&=& a\sqrt{\mu}(Z_{d}Q^{-1/2}-Z_{d}^{2/3}) + b + c\sqrt{l(l+1)} \nonumber \\ &&+ d\sqrt{I(I+1)} +h
\end{eqnarray}
where the half-life is in the unit of seconds and $I=(N-Z)/A$ represents the isospin of the parent nucleus. The term $\sqrt{l(l+1)}$ reflects the hindrance effect of the centrifugal barrier. The last term 'h' is the blocking effect of unpaired nucleon which is taken as zero for even-even nuclei. Other symbols a, b, c, d, and h (for odd nucleons) represent the parameters to be determined in the least square fit to the experimental data. This formula is a new modified version of the TN formula and therefore named as NMTN formula, hereafter. We have fitted the NMTN formula for two regions on either side of $N=126$ using the data-set of a total of 349 nuclei, 174 for $N\leq126$ and 175 for $N>126$, taken from NUBASE2020 \cite{audi2020}. To check the validity and efficiency of the several fittings and other similar formulas on 349 data, we calculate RMSE as given by Eqn (1). First, the NMTN formula is fitted with only 2 terms (with a and b coefficients) then, systematically, by adding $\sqrt{l(l+1)}$, $\sqrt{I(I+1)}$ and the last terms which result to the step-down of RMSE values from 0.63 to 0.43, 0.42 and, 0.38, respectively, for the full dataset of 349 nuclei. The coefficients of the NMTN formula with the least RMSE are mentioned in Table \ref{coeff} for both regions i.e. $N\leq126$ and $N>126$, along with the combined RMSE value (0.38) between estimated half-lives and experimental half-lives. This formula is also tested on the same trail of k-cross validation and found with remarkable efficiency as mentioned in Table \ref{alpha-rmse} over all the considered formulas. \par
\begin{figure}[!htbp]
\centering
\includegraphics[width=10.2cm, height=7.6cm]{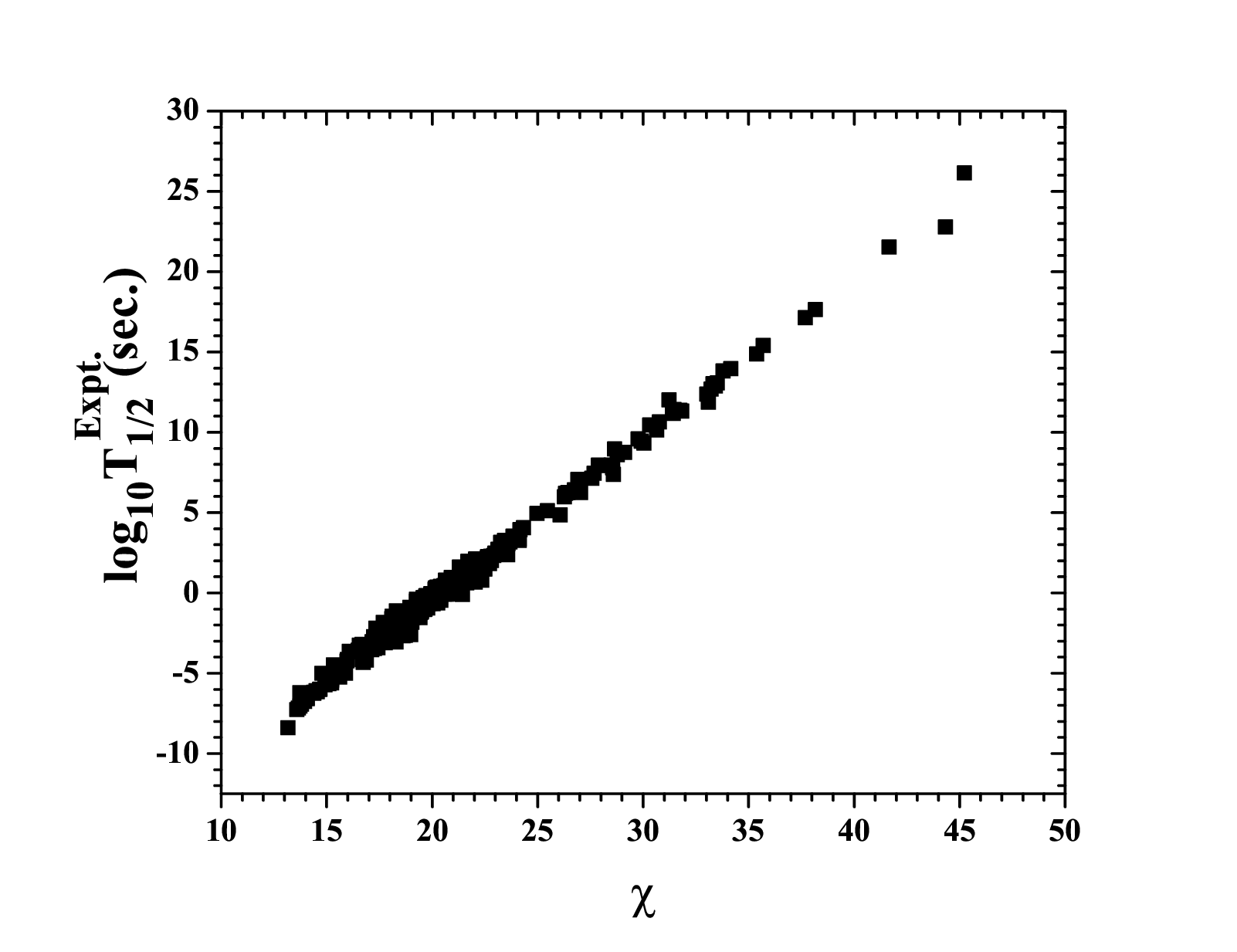}
\caption{Logarithm of $\alpha$-decay half-lives for ($log_{10}T_{1/2}^{Expt.}$) of 349 nuclei \cite{audi2020} as a function of NMTN formula (see the text for details).}
\label{Fig:III-VSS}
\end{figure}

\begin{table*}[!htbp]
\caption{The coefficients of NMTN formula for $\alpha$-decay half-life and modified Deng and Zhang (MDZ) formula for $\alpha$-particle preformation factor. In the last column root mean square error (RMSE) between estimated values and experimental(extracted) values for 349 nuclei are also shown.}
\centering
\resizebox{1.0\textwidth}{!}{%
\begin{tabular}{l@{\hskip 0.4in}|c@{\hskip 0.4in}c@{\hskip 0.4in}c@{\hskip 0.4in}c@{\hskip 0.4in}c@{\hskip 0.4in}c@{\hskip 0.4in}c@{\hskip 0.4in}c@{\hskip 0.4in}|c}
\hline
\hline
Formula  & Region  & Nuclei   &    a      &   b      &     c   &      d    &   e     &   h    & $RMSE$\\
\hline
NMTN       &            & E-E  &         &           &          &          &         &0.0000  & 0.38 \\
Formula    & N$\leq$126 & O-A  & 0.7838  & -20.4118  & 0.1630   &  2.9908  & -       &0.1562 &      \\
           &            & O-O  &         &           &          &          &         &0.3515  &      \\
\cline{2-9}
           &            & E-E  &         &           &          &          &         &0.0000  &      \\
           & N$>$126    & O-A  &0.7988   & -18.3048  & 0.3041   & -3.0399  & -       &0.1198  &      \\
           &            & O-O  &         &           &          &          &         &0.0986 &      \\
\hline
MDZ        &            & E-E  &         &           &          &          &         &0.0000 & 0.29 \\
Formula    & N$\leq$126 & O-A  &16.8340  & -4.9133   &  0.0064  &  -0.0848 & -1.2591 &-0.1243 &      \\
           &            & O-O  &         &           &          &          &         &-0.6137&      \\
\cline{2-9}
           &            & E-E  &         &           &          &          &         &0.0000 &      \\
           & N$>$126    & O-A  & 34.0926 & -9.5946   & 0.0345   & -0.1917  &-2.5603  &-0.1449&      \\
           &            & O-O  &         &           &          &          &         &-0.5081&      \\
\hline
\hline
\end{tabular}}
\label{coeff}
\end{table*}
Mathematically, lower values of RMSE may be attributed to the number of free parameters in the present formula. To justify the applicability of our proposed formula and the increased number of parameters due to partition at $N=126$, we have calculated $\chi^2$ per degree of freedom (defined in Eqn. (2)) for all the considered formulas from Table \ref{alpha-rmse}. In Eqn. (2), $N_d$ and $N_p$ refer to the total data points and the total number of parameters used for the fitting, respectively. It is gratifying to note from Table \ref{alpha-rmse} that the value of $\chi^2$ per degree of freedom is found minimum for the proposed NMTN formula when compared to the other formula with their corresponding $\chi^2$ which directly correlated to the number of parameters used. For instance, the NMTN formula results in $\chi^2$=0.15 with a total of 12 parameters whereas the nearest competitor formula i.e. Soylu 2021 results $\chi^2$=0.61 with only 6 parameters. It is important to emphasize here that all the considered formulas are refitted on the same dataset of 349 nuclei (without breaking in the partition at $N=126$) to preserve the equivalency among formulas and to keep the total number of parameters as such. The quality of accuracy and consistency of proposed NMTN formula can be verified by Viola-Seaborg type plot in which experimental half-lives of 349 nuclei are plotted as a function of $\chi$= a$\sqrt{\mu}(Z_{d}Q^{-1/2}-Z_{d}^{2/3})$+ c$\sqrt{l(l+1)}$ + d$\sqrt{I(I+1)}$ +h in Fig. \ref{Fig:III-VSS}.
Hence, the proposed NMTN formula, accumulating various crucial microscopic properties, with a total of 12 coefficients (6 for N$\leq$126 + 6 for N$>$126) qualifies for a higher degree of accuracy and can be utilized to estimate $\alpha$-decay half-life more effectively and meticulously. Since, superheavy nuclei have much greater experimental uncertainties associated with measured half-lives than the rest of the nuclear chart (even an order of magnitude greater, in some cases), therefore the inclusion of all available uncertainties in the data set of the NMTN formula leads to $\pm$0.3821 and $\pm$0.6324 average uncertainty in all the prediction of $\alpha$-decay half-lives of the present article for N$\leq$126 and for N$>$126 nuclei, respectively.\par

\subsection{Formula for $\alpha$-particle preformation factor}
 Another quantity that is found to be crucial for the $\alpha$-decay is the $\alpha$-particle preformation factor, which represents the probability of an $\alpha$-cluster formation on the surface of the decaying parent nucleus \cite{seif2015,zhang2009,guo2015,dengplb2021}. However, describing the formation of $\alpha$–particles on the surface of an atomic nucleus from two protons and two neutrons remains a considerable theoretical challenge \cite{delion1992,delion1994,delion2000,lane1958}. For a simplified picture, the expression of the preformation factor can be extracted from the following equation of experimental $\alpha$-decay constant $\lambda_{Expt.}$:
\begin{equation}
\lambda_{Expt.} = \frac{ln 2}{T_{Expt.}^{1/2}} = P_{\alpha} \nu P
\end{equation}
where $T_{Expt.}^{1/2}$ is the $\alpha$-decay half-life and $P_{\alpha}$ is the preformation factor. $\nu$ is the assault frequency which is expressed as:
\begin{equation}
\nu = \frac{1}{2R_0}\sqrt{\frac{2E_{\alpha}}{M_{\alpha}}}
\end{equation}
where R$_0$ denotes the radius of the $\alpha$ decay parent nucleus (index $i$ = 0) and can be obtained by following formula which can also be used for the calculations of radii of fragments (index $i$ = 1, 2)
\begin{equation}
R_i = 1.28 A_{i}^{1/3} - 0.76 + 0.8A_{i}^{-1/3} \hspace{0.5cm} (i = 0, 1, 2).
\end{equation}
$E_{\alpha}=\frac{A-4}{A}Q_{\alpha}$ is the kinetic energy of the $\alpha$-particle, with A and $Q_{\alpha}$ being the mass number and $\alpha$ decay energy of the parent nucleus. $M_{\alpha}$ represents the mass of the $\alpha$ particle.

In Eqn. (4), P is the barrier penetrating probability which can be expressed by the following equation using the Wentzel-Kramers-Brillouin (WKB) approximation:
\begin{equation}
P = exp \left[-\frac{2}{\hbar}\int_{r_{in}}^{r_{out}}\sqrt{2B(r)[E_{r}-E(sphere)]}dr\right]
\end{equation}
where $r$ is the center-of-mass distance between the $\alpha$ cluster and daughter nucleus. The classical turning points $r_{in}$ and $r_{out}$ satisfy the conditions $r_{in}$ = R$_1$ + R$_2$ and E($r_{out}$) = $Q_{\alpha}$. B(r) = $\mu$ denotes the reduced mass between the $\alpha$ particle and the daughter nucleus.\\

Likewise as Eqn. (4), we can define theoretical $\alpha$-decay constant $\lambda_{Th}$ as:
\begin{equation}
\lambda_{Th} = \frac{ln 2}{T_{Th}^{1/2}} = P_{0} \nu P
\end{equation}
with $P_{0}$=1 in principle. From these equations, the experimental $\alpha$-particle preformation factor along with its logarithmic form can be extracted as:

  \begin{equation}
  P_{\alpha}^{Expt.} = \frac{\lambda_{Expt.}}{\lambda_{Th}} =  \frac{T_{Th}^{1/2}}{T_{Expt.}^{1/2}}
  \end{equation}

  \begin{equation}
 log_{10} P_{\alpha}^{Expt.} = log_{10}T_{Th}^{1/2}-log_{10}T_{Expt.}^{1/2}
  \end{equation}


\begin{figure}[!htbp]
\centering
\includegraphics[width=0.55\textwidth]{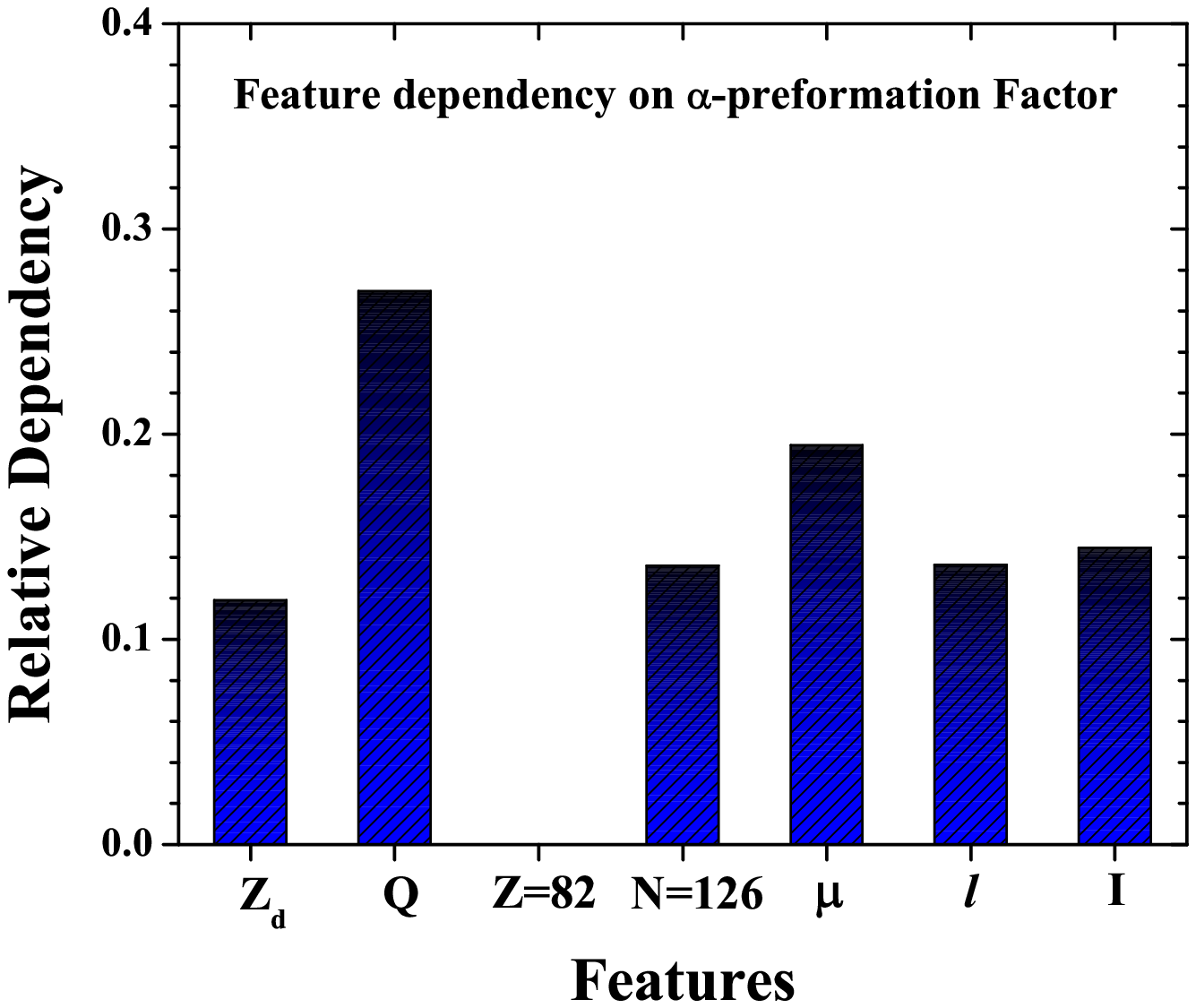}
\caption{Relative dependencies of $\alpha$-particle preformation factors on various quantities (features).}
\label{Fig:III}
\end{figure}

\begin{table*}[!htbp]
\caption{Root mean square errors (RMSE) for various formulas of $\alpha$-particle preformation factor using different sets of data for fitting and testing (see the text for details). The second last column contains the average value of RMSE for all 5 sets. The last column lists the value of $\chi^2$ per degree of freedom computed by using the total number of fitted coefficients for the respective formula.}
\centering
\resizebox{1.0\textwidth}{!}{%
\begin{tabular}{l@{\hskip 0.5in}|c@{\hskip 0.5in}c@{\hskip 0.5in}c@{\hskip 0.5in}c@{\hskip 0.5in}c|@{\hskip 0.5in}c@{\hskip 0.5in}|c}
\hline
\hline
\multicolumn{1}{l}{Formula for}&\multicolumn{6}{|c|}{RMSE}&\multicolumn{1}{c}{$\chi^2$}\\
\cline{2-7}
$\alpha$-particle preformation factor       &  set 1 & set 2 & set 3 & set 4 & set 5& Average & \\
\hline
MDZ 2023 (present work)                       & 0.4020  &0.3591	  &0.4090	        &0.3766          & 0.6315          	& 0.4357               &   0.19  \\
DZ 2021 \cite{dengplb2021}                    &  0.3151	&0.4938	  &0.3716        	&0.3684	         & 0.6044          	& 0.4596                     &  0.21     \\
DZ 2020 \cite{deng2020}                       & 0.3608	&	0.6402&	  0.4327        &	0.4290       &    0.6348       	&    0.4657              &  0.28     \\
SJ 2021 \cite{santhosh2021}                   & 0.3839	&	0.6450&	  0.3547        &	0.3692       &    0.7137        &    0.4933            &  0.28    \\
SN 2018  \cite{santhosh2018}                  & 0.4345	&	0.4400&   0.3790    	&	   0.3844    &       0.7424     &       0.5003      &  0.25        \\
ZR 2008 \cite{royer2008Preformation}          & 0.3713	&	0.6491&   0.3547  	    &   0.3799       &	  0.7116        &    0.5217           &  0.28    \\
YI 2022 \cite{yahya2022}                     & 0.5277	&	0.5678&	  0.3773        &	0.3864       &	   0.8243       &     0.5367        &  0.32        \\

\hline
\hline
\end{tabular}}
\label{p-rmse}
\end{table*}

Recently, Deng \textit{et al.} \cite{dengplb2021} have shown a correlation between $\alpha$-particle preformation factors and $\alpha$-decay energies and proposed a new formula in which logarithmic value of extracted experimental $\alpha$-particle preformation factor is linearly dependent on the reciprocal of the square root of experimental $\alpha$-decay energy Q$_{\alpha}$$^{-1/2}$. We have also worked on the dependency of $\alpha$-particle preformation factors on various quantities implied by the empirical formulas of $\alpha$-decay viz. proton number of daughter nucleus ($Z_d$), Q-value, shell effects of Z$=$82 and N$=$126, reduced mass ($\mu$), angular momentum ($l$), and nucleus asymmetry ($I$). We estimate the importance of the above-mentioned quantities (features) for a predictive modeling problem using gradient boosting by applying the XGBoost library in Python \cite{chen2016xgboost}. This method is straightforward to retrieve the importance of each feature in terms of a score that indicates how valuable each feature is in the construction of the boosted decision trees within the model. The more a feature is used to make key decisions with a trained model, the higher its relative importance \cite{rajbahadur2022} will be. This feature importance is calculated explicitly for each feature in the dataset, allowing features to be ranked and compared to each other which is shown in Fig. \ref{Fig:III} for the above-mentioned features (quantities) for the $\alpha$-preformation factor.\par

As can be seen from Fig. \ref{Fig:III} the preformation factor relies infinitesimally on the shell effect of Z$=$82 compared to that of N$=$126 which is in accord with the preceding analysis of $\alpha$-decay and can also be seen in Fig. \ref{Fig:126} (iii) for the experimental preformation factor. Additionally, the Q-values dependence similar to the Ref. \cite{dengplb2021} can indeed be affirmed by this analysis from Fig. \ref{Fig:III}. As a result, the formula of the preformation factor given by Deng and Zhang \cite{dengplb2021} accommodates most of these dependencies viz. shell effect of N$=$126, $Q$-value, $l$-value, and odd-even effect. However, the formula still misses one of the crucial dependencies on the asymmetry term (isospin term) that is found fairly significant in Fig. \ref{Fig:III} as well as by the analysis of $\alpha$-decay in the previous subsection. Hence, we add the asymmetry term ($\sqrt{I(I+1)}$) to the Deng and Zhang formula (DZ) \cite{dengplb2021} and refit it. In view of these perspectives, the modified empirical formula (MDZ) to calculate $\alpha$-particle preformation factor shapes as:

\begin{eqnarray}\label{preformation-factor}
 log_{10}P_{\alpha} &=& a+ b (A_{\alpha}^{1/6}+A_{d}^{1/6})+c\frac{N}{\sqrt{Q_{\alpha}}} + d\sqrt{l(l+1)} \nonumber \\ &&+e\sqrt{I(I+1)}+h
\label{mdz}
\end{eqnarray}

here $A_{\alpha}$ and $A_{d}$ are the mass numbers of the $\alpha$-particle and daughter nucleus. Whereas, $N$ and $Q_{\alpha}$ represent neutron number and $\alpha$-decay  energy of the parent nucleus. To demonstrate the accuracy level and the validity of this particular MDZ formula, we have compared it with some similar other empirical formulas \cite{deng2020,dengplb2021,santhosh2021,santhosh2018,royer2008Preformation,yahya2022} of preformation factor in the same way as that applied for the formulas of $\alpha$-decay half-lives. The results of various $\alpha$-particle preformation formulas are mentioned in Table \ref{p-rmse} from where, it is gratifying to note that the MDZ formula performs better than the other formulas \cite{deng2020,dengplb2021,santhosh2021,santhosh2018,royer2008Preformation,yahya2022} even on k-cross validation, and hence will be employed for the estimation of the preformation factor.\par
\begin{figure}[!htbp]
\centering
\includegraphics[width=0.5\textwidth]{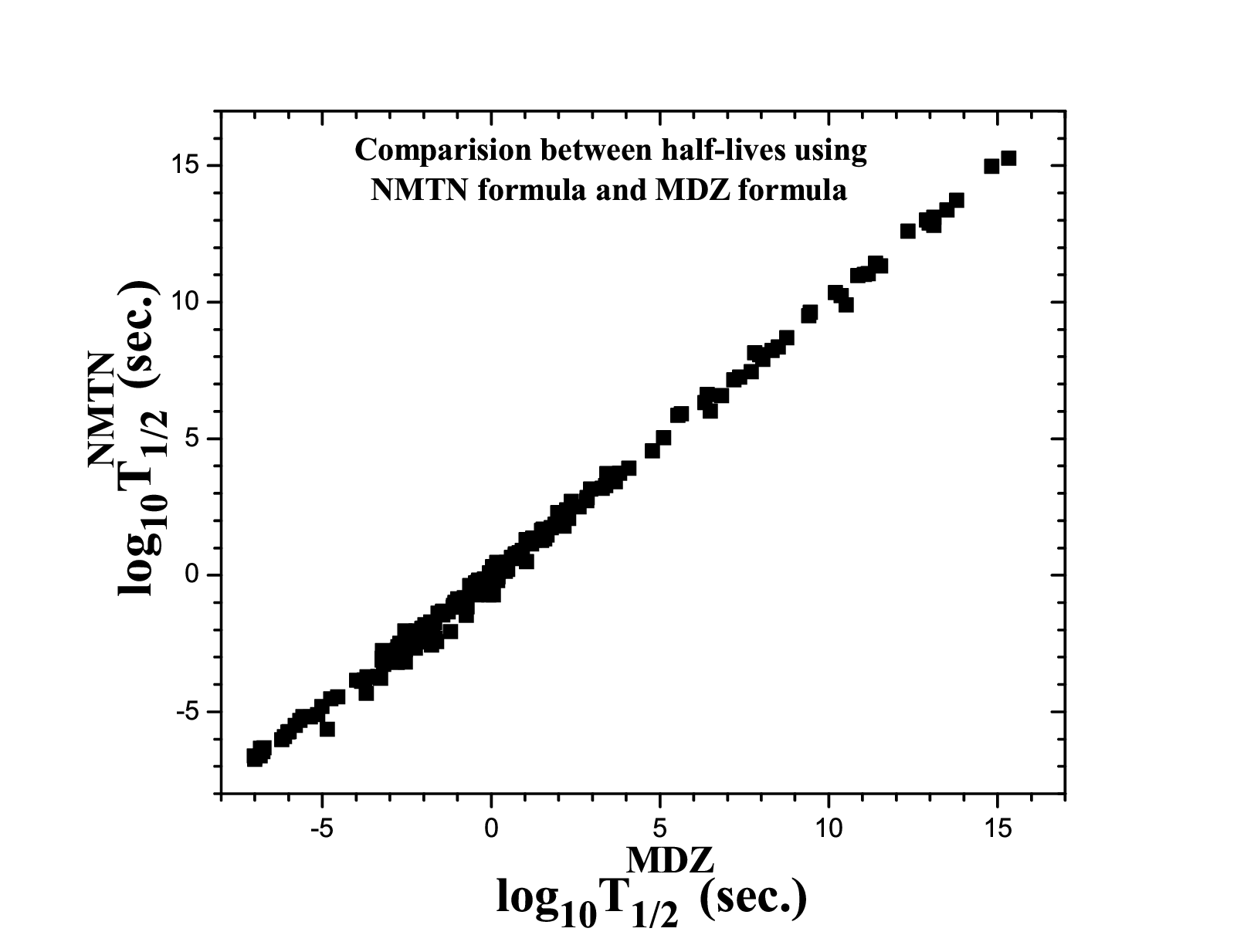} \caption{Logarithm of $\alpha$-decay half-lives calculated by using NMTN formula compared with the half-lives calculated by using MDZ formula.}
\label{Fig:comparision}
\end{figure}

The fitting coefficients are obtained by considering the nuclei with N$\le$126 and N$>$126 and are mentioned in Table \ref{coeff} along with the RMSE on all data-set which comes up with significantly lower value i.e. 0.29.
It is important to point out here that Eqn. \ref{mdz} is modified for the same data of experimental $\alpha$-particle preformation factors as used in the Refs. \cite{deng2020,dengplb2021}. This Eqn. \ref{mdz} along with coefficients in Table \ref{coeff} can also be used to calculate $\alpha$-decay half-lives (please see the last column of Table III of Ref. \cite{deng2020}). The half-lives calculated by using NMTN formula (Eqn. \ref{NTN}) and MDZ formula (Eqn. \ref{mdz}) are found very close to each other. A comparison for these half-lives can be found in the Fig. \ref{Fig:comparision}.
\section{Estimation of half-lives and preformation factor}
The previous analysis has offered two effective, precise, and modified formulas based on crucial physical aspects of decay to evaluate half-lives and the preformation factor of $\alpha$-decay. These formulas can be utilized to estimate the half-lives and the preformation factor for the unknown region of the periodic chart, especially towards the northeast corner. With this in view, we have listed the half-lives and preformation factor for a few selected nuclei from NUBASE2020 \cite{audi2020} for which Q-values and half-lives are still not measured. These nuclei are included in NUBASE2020 \cite{audi2020} with the non-experimental value of Q, half-life, spin, and parity estimated from Trends in Neighboring Nuclei (TNN). The decay modes for these nuclei are still undetermined. These unknown nuclei and their properties are listed in Table \ref{prediction} with the symbol $\#$: indicating estimated data. Spin and parity for these nuclei are also taken from Ref. \cite{mollerparity} (mentioned by $\ast$), if unavailable in NUBASE2020.
Our predicted values of half-lives using the NMTN formula are very close to those estimated by using TNN in NUBASE2020 \cite{audi2020}. Additionally, the respective preformation factors calculated by using the MDZ formula are mentioned. As a purpose of completeness, we have mentioned the calculated values of half-lives of $\alpha$-decay by using the NTMN formula for the whole 349 known data along with other details in the supplementary material ($\textit{supplimentry-material-1-349-data.dat}$). \par

\begin{table*}[!htbp]
\caption{$\alpha$-decay half-lives from NMTN formula and the $\alpha$-decay preformation factors from MDZ formula for few selective unknown nuclei from NUBASE2020 \cite{audi2020}. Q-value, half-life, spin and parity are estimated from Trends in Neighboring Nuclei (TNN) in the NUBASE2020 as indicated by $\#$ symbol. Spin and parity of nuclei mentioned by $\ast$ are taken from Ref. \cite{mollerparity} if these are not available in NUBASE2020. $l$ is the angular momentum of the emitted $\alpha$-particle which is calculated by standard selection rules \cite{denisov2009}.}
\centering
\def\arraystretch{1.15}
\resizebox{1.00\textwidth}{!}{%
{\begin{tabular}{cccccccc|cccccccc}
 \hline
\multicolumn{1}{c}{$\alpha$-}&
\multicolumn{4}{c}{NUBASE2020-TNN($\#$)}&
\multicolumn{1}{c}{$l$}&
 \multicolumn{2}{c|}{Present Work}&
\multicolumn{1}{c}{$\alpha$-}&
\multicolumn{4}{c}{NUBASE2020-TNN($\#$)}&
\multicolumn{1}{c}{$l$}&
 \multicolumn{2}{c}{Present Work}\\
\cline{2-5} \cline{7-8}\cline{10-13}\cline{15-16}
\multicolumn{1}{c}{Transition}&
\multicolumn{1}{c}{Q}& \multicolumn{1}{c}{log$_{10}$ T$_{\alpha}$}&\multicolumn{1}{c}{j$_{p}$}& \multicolumn{1}{c}{j$_{d}$}& \multicolumn{1}{c}{}& \multicolumn{1}{c}{log$_{10}$ T$_{\alpha}$}&\multicolumn{1}{c|}{P$\alpha$}&
\multicolumn{1}{c}{Transition}&
\multicolumn{1}{c}{Q}& \multicolumn{1}{c}{log$_{10}$ T$_{\alpha}$}&\multicolumn{1}{c}{j$_{p}$}& \multicolumn{1}{c}{j$_{d}$}& \multicolumn{1}{c}{}& \multicolumn{1}{c}{log$_{10}$ T$_{\alpha}$}&\multicolumn{1}{c}{P$\alpha$}\\
\multicolumn{1}{c}{}&
\multicolumn{1}{c}{(MeV)}& \multicolumn{1}{c}{(sec.)}&\multicolumn{1}{c}{}& \multicolumn{1}{c}{}& \multicolumn{1}{c}{}& \multicolumn{1}{c}{(NMTN)}&\multicolumn{1}{c|}{(MDZ)}&
\multicolumn{1}{c}{}&
\multicolumn{1}{c}{(MeV)}& \multicolumn{1}{c}{(sec.)}&\multicolumn{1}{c}{}& \multicolumn{1}{c}{}& \multicolumn{1}{c}{}& \multicolumn{1}{c}{(NMTN)}&\multicolumn{1}{c}{(MDZ)}\\

 \hline
$^{107}$I $\rightarrow$$^{103}$Sb &  4.82 &-4.70 &5/2$^{+}$ & 5/2$^{+}$ & 0 & -5.39 &0.0825& $^{271}$Hs$\rightarrow$$^{267}$Sg& 9.46 & 1.00  &$^\ast$13/2$^{-}$  &$^\ast$3/2$^{+}$ & 5 & 1.11 &0.0011 \\
$^{169}$Au$\rightarrow$$^{165}$Ir &  7.38 &-3.82 &1/2$^{+}$ & 1/2$^{+}$ & 0 & -3.60 &0.0781& $^{274}$Hs$\rightarrow$$^{270}$Sg& 9.55 & -0.30 &0$^{+}$     &0$^{+}$   & 0 & -0.97              &0.0148 \\
$^{220}$U $\rightarrow$$^{216}$Th &  10.29&-7.22 &0$^{+}$   & 0$^{+}$   & 0 & -7.01 &0.0580& $^{276}$Hs$\rightarrow$$^{272}$Sg& 9.24 & -1.00 &0$^{+}$     &0$^{+}$   & 0 & -0.09              &0.0149 \\
$^{221}$Np$\rightarrow$$^{217}$Pa &  10.43&-7.52 &9/2$^{-}$ & 9/2$^{-}$ & 0 & -6.91 &0.0409& $^{265}$Mt$\rightarrow$$^{261}$Bh& 11.12& -2.70 &$^\ast$11/2$^{+}$  &5/2$^{-}$ & 3 & -3.38       &0.0024\\
$^{224}$Pu$\rightarrow$$^{220}$U  &  9.84 &-5.00 &0$^{+}$   & 0$^{+}$   & 0 & -5.45 &0.0570& $^{267}$Mt$\rightarrow$$^{263}$Bh& 10.87& -2.00 &$^\ast$11/2$^{+}$  &5/2$^{-}$ & 3 & -2.83       &0.0024\\
$^{225}$Pu$\rightarrow$$^{221}$U  &  9.36 &-4.00 &7/2$^{+}$ & 9/2$^{+}$ & 2 & -3.43 &0.0143& $^{269}$Mt$\rightarrow$$^{265}$Bh& 10.48& -1.00 &$^\ast$11/2$^{+}$  &5/2$^{-}$ & 3 & -1.90       &0.0023\\
$^{227}$Pu$\rightarrow$$^{223}$U  &  8.3  &0.30  &5/2$^{+}$ & 7/2$^{+}$ & 2 & -0.51 &0.0162& $^{271}$Mt$\rightarrow$$^{267}$Bh& 9.91 & -0.40 &$^\ast$11/2$^{+}$  &5/2$^{-}$ & 3 & -0.44       &0.0024\\
$^{231}$Am$\rightarrow$$^{227}$Np &  7.41 &1.78  &5/2$^{-}$ & 5/2$^{+}$ & 1 & 2.45  &0.0280& $^{279}$Mt$\rightarrow$$^{275}$Bh& 9.38 & 1.30  &$^\ast$11/2$^{+}$  &5/2$^{-}$ & 3 & 0.96        &0.0021\\
$^{231}$Cm$\rightarrow$$^{227}$Pu &  8.08 &1.30  &3/2$^{+}$ & 5/2$^{+}$ & 2 & 0.84  &0.0157& $^{283}$Ds$\rightarrow$$^{279}$Hs& 8.91 & 1.78  &$^\ast$5/2$^{+}$   &$^\ast$3/2$^{+}$ & 2 & 2.39 &0.0034 \\
$^{232}$Cm$\rightarrow$$^{228}$Pu &  7.8  &1.00  &0$^{+}$   & 0$^{+}$   & 0 & 0.89  &0.0662& $^{284}$Ds$\rightarrow$$^{280}$Hs& 8.62 & 1.78  &0$^{+}$     &0$^{+}$   & 0 & 2.47               &0.0146 \\
$^{235}$Bk$\rightarrow$$^{231}$Am &  7.94 &1.78  &3/2$^{-}$ & 5/2$^{-}$ & 2 & 1.61  &0.0143& $^{273}$Rg$\rightarrow$$^{269}$Mt& 11.16& -2.70 &$^\ast$3/2$^{-}$   &$^\ast$13/2$^{+}$& 5 &-2.39 &0.0008 \\
$^{239}$Es$\rightarrow$$^{235}$Bk &  8.44 &0.00  &3/2$^{-}$ & 3/2$^{-}$ & 0 & -0.10 &0.0348& $^{283}$Rg$\rightarrow$$^{279}$Mt& 9.37 & 2.08  &$^\ast$13/2$^{+}$  &$^\ast$11/2$^{+}$& 2 & 1.29 &0.0031 \\
$^{263}$Lr$\rightarrow$$^{259}$Md &  7.68 &4.26  &1/2$^{-}$ & 7/2$^{-}$ & 4 & 5.01  &0.0029& $^{280}$Cn$\rightarrow$$^{276}$Ds& 10.69& -2.30 &0$^{+}$     &0$^{+}$   & 0 & -2.85              &0.0110 \\
$^{264}$Rf$\rightarrow$$^{260}$No &  8.04 &3.56  &0$^{+}$   & 0$^{+}$   & 0 & 2.57  &0.0261& $^{287}$Cn$\rightarrow$$^{283}$Ds& 9.12 & 1.48  &$^\ast$1/2$^{+}$   &$^\ast$5/2$^{+}$ & 2 & 2.35 &0.0030 \\
$^{266}$Rf$\rightarrow$$^{262}$No &  7.61 &4.16  &0$^{+}$   & 0$^{+}$   & 0 & 4.14  &0.0277& $^{288}$Cn$\rightarrow$$^{284}$Ds& 9.05 & 1.00  &0$^{+}$     &0$^{+}$   & 0 & 1.70               &0.0124 \\
$^{265}$Db$\rightarrow$$^{261}$Lr &  8.4  &2.95  &9/2$^{+}$ & 1/2$^{-}$ & 5 & 3.47  &0.0015& $^{287}$Nh$\rightarrow$$^{283}$Rg& 9.65 & 1.30  &$^\ast$7/2$^{-}$   &$^\ast$13/2$^{+}$& 3 & 1.37 &0.0018 \\
$^{269}$Db$\rightarrow$$^{265}$Lr &  8.49 &4.03  &9/2$^{+}$ & 1/2$^{-}$ & 5 & 3.12  &0.0013& $^{291}$Fl$\rightarrow$$^{287}$Cn& 9.71 & 1.00  &$^\ast$1/2$^{+}$   &$^\ast$1/2$^{+}$ & 0 & 0.42 &0.0073 \\
$^{268}$Sg$\rightarrow$$^{264}$Rf &  8.3  &2.08  &0$^{+}$   & 0$^{+}$   & 0 & 2.34  &0.0228& $^{291}$Mc$\rightarrow$$^{287}$Nh& 10.3 & 0.00  &$^\ast$5/2$^{-}$   &$^\ast$7/2$^{-}$ & 2 &-0.19 &0.0023 \\
$^{263}$Bh$\rightarrow$$^{259}$Db &  10.08&-0.70 &5/2$^{-}$ & 9/2$^{+}$ & 3 & -1.41 &0.0029& $^{289}$Lv$\rightarrow$$^{285}$Fl& 11.1 & -1.80 &$^\ast$5/2$^{+}$   &$^\ast$3/2$^{+}$ & 2 &-1.93 &0.0021 \\
$^{269}$Bh$\rightarrow$$^{265}$Db &  8.67 &1.78  &5/2$^{-}$ & 9/2$^{+}$ & 3 & 2.61  &0.0032& $^{291}$Ts$\rightarrow$$^{287}$Mc& 11.48& -2.70 &$^\ast$3/2$^{-}$   &$^\ast$5/2$^{-}$ & 2 &-2.58 &0.0019 \\
$^{273}$Bh$\rightarrow$$^{269}$Db &  9.11 &1.78  &1/2$^{-}$ & 9/2$^{+}$ & 5 & 1.79  &0.0010& $^{293}$Og$\rightarrow$$^{289}$Lv& 11.92& -3.00 &$^\ast$1/2$^{+}$   &$^\ast$5/2$^{+}$ & 2 &-3.34 &0.0017 \\
\hline
\end{tabular}}
}
\label{prediction}
\end{table*}

\begin{figure}[!htbp]
\centering
\includegraphics[width=0.5\textwidth]{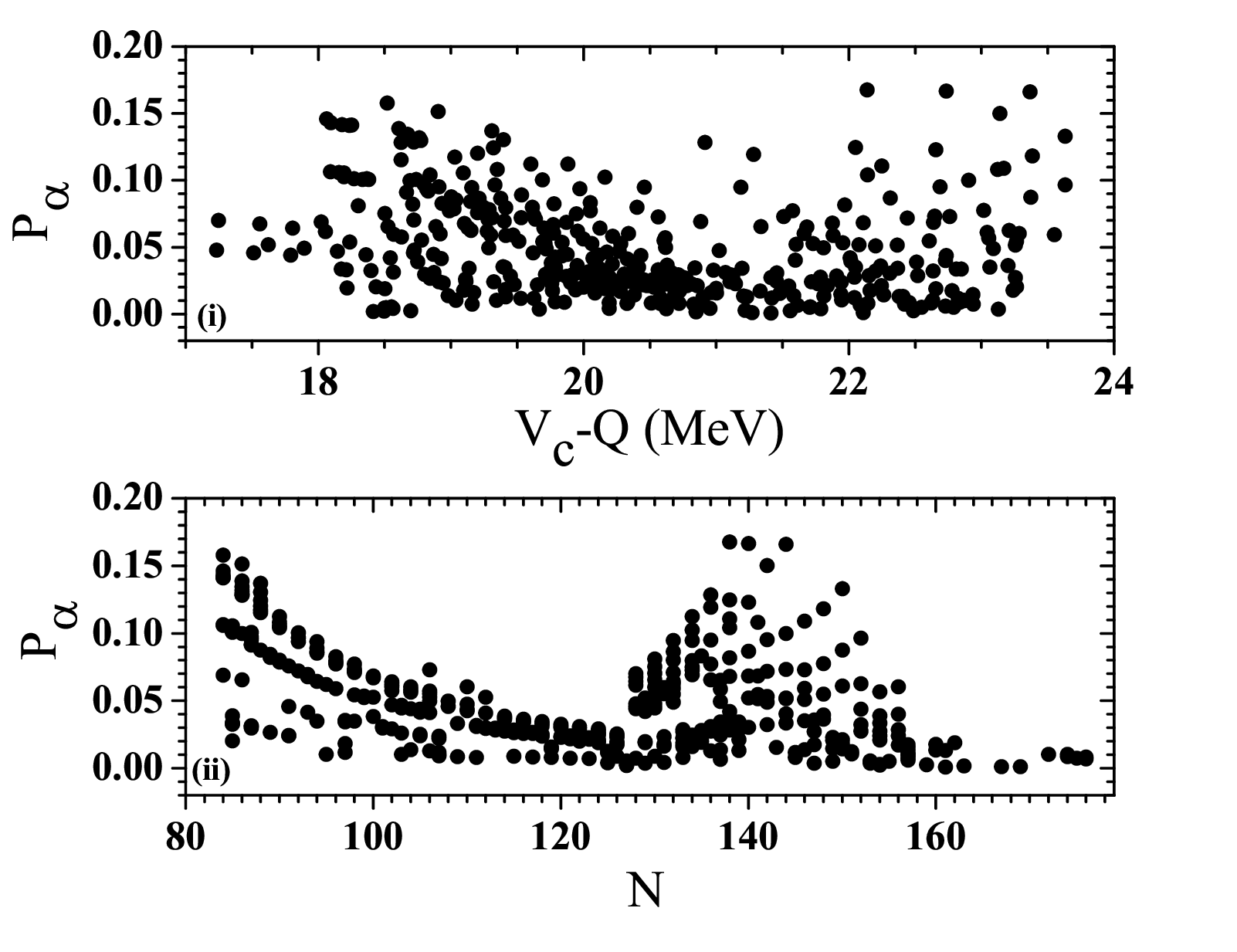}\caption{Predicted $\alpha$-preformation factor  by using MDZ formula as a function of (i) fragmentation potential and (ii) neutron number N.}
\label{Fig:p-fp-n}
\end{figure}

The variation of the preformation factor for the whole periodic chart can also be reckoned with Fig. \ref{Fig:p-fp-n} where calculated values of P$_{\alpha}$ are plotted as a function of fragmentation potential \cite{delion2009}. The fragmentation potential can be calculated as the difference between the top of the Coulomb barrier and the respective $Q$-value of the nuclei. (for more details one can refer to Eqn. (3.4) of Ref. \cite{delion2009}).
\begin{equation}
    V_{frag} (r_b) =  V_C (r_b) -Q
\end{equation}
here Coulomb barrier is calculated by using $V_C = Z_c Z_d e^{2}/{r_b} $ with the touching radius $r_b=1.2 (A_c^{1/3} + A_d^{1/3})$. $Z_c, Z_d$, and $A_c, A_d$ are the proton numbers and mass numbers of cluster ($\alpha$) and daughter nucleus, respectively.

The convergent behavior of the preformation factor at N=126 can be seen from Fig. \ref{Fig:p-fp-n} (ii) which corresponds to the region with fragmentation potential $\sim$21 MeV from Fig. \ref{Fig:p-fp-n} (i). This kind of trend is in line with the odd clustering behaviour at N = 126, reported recently \cite{delion2020}.

\begin{table*}[!htbp]
\caption{Estimation of $\alpha$-decay half-lives for the most recent (probable) measurements of $\alpha$-decay chains. The data are taken from respective references mentioned in front of the nucleus itself, otherwise the $Q$-values (mentioned by $^{\dag}$ superscript) are taken from Ref. \cite{ws42014} and j$^{\pi}$ values for parent and daughter nuclei are taken from NUBASE2020 \cite{audi2020}. The first column index (n) is for labeling the data (see text for details).}
\centering
\resizebox{0.9\textwidth}{!}{%
\begin{tabular}{c@{\hskip 0.2in}|c@{\hskip 0.2in}|c@{\hskip 0.7in}c@{\hskip 0.7in}c@{\hskip 0.7in}c@{\hskip 0.7in}c@{\hskip 0.7in}c}
\hline
Index&Nucleus&Q&log$_{10}$ T$_{\alpha}$& j$_{p}$& j$_{d}$&$l$&\multicolumn{1}{c}{Present Work}\\
(n)&&(MeV)&(sec.)& & &&log$_{10}$ T$_{\alpha}(sec.)$\\
\hline
\multicolumn{8}{c}{$^{112}$Ba$\rightarrow$$^{108}$Xe$\rightarrow$$^{104}$Te$\rightarrow$$^{100}$Sn}\\
\hline
1&$^{112}$Ba                    & 3.967$^{\dag}$     &         -                   &  0$^{+}$  & 0$^{+}$  & 0   & -0.67   \\
2&$^{108}$Xe \cite{auranen2018} & 4.4     &  -4.24$^{+0.08}_{-0.18}$        &  0$^{+}$  & 0$^{+}$  & 0  & -3.71   \\
3&$^{104}$Te \cite{auranen2018} & 4.9    &   $<$-7.74                  &  0$^{+}$  & 0$^{+}$  & 0   & -6.55  \\
\hline
\multicolumn{8}{c}{$^{208}$Pa$\rightarrow$$^{204}$Ac$\rightarrow$$^{200}$Fr$\rightarrow$$^{196}$At}\\
\hline
4&$^{208}$Pa                  & 8.665$^{\dag}$ &   -                      &  3$^{+}$  & 3$^{+}$ & 0  & -2.95     \\
5&$^{204}$Ac \cite{huang2022} & 7.948          &  -2.13$^{+0.88}_{-0.08}$ &  3$^{+}$  & 3$^{+}$ & 0  & -1.50      \\
6&$^{200}$Fr \cite{huang2022} & 7.48           &  -1.12$^{+0.37}_{-0.09}$ &  3$^{+}$  & 3$^{+}$ & 0  & -0.68      \\
7&$^{196}$At \cite{huang2022} & 7.054          &  -0.50$^{+0.06}_{-0.08}$ &  3$^{+}$  & 3$^{+}$ & 0  & 0.06       \\
\hline
\multicolumn{8}{c}{$^{298}$Og$\rightarrow$$^{294}$Lv$\rightarrow$$^{290}$Fl$\rightarrow$$^{286}$Cn$\rightarrow$$^{282}$Ds$\rightarrow$$^{278}$Hs}\\
\hline
8&$^{298}$Og                    & 12.187$^{\dag}$   &         -                   &  0$^{+}$  & 0$^{+}$  & 0  &  -4.84  \\
9&$^{294}$Lv                    & 10.669$^{\dag}$   &         -                   &  0$^{+}$  & 0$^{+}$  & 0  &  -1.77  \\
10&$^{290}$Fl \cite{hofmann2016} &  9.846  &    1.32$^{+1.13}_{-0.22}$        &  0$^{+}$  & 0$^{+}$  & 0  &  -0.09         \\
11&$^{286}$Cn \cite{hofmann2016} &  8.793  &    2.81$^{+1.88}_{-0.22}$     &  0$^{+}$  & 0$^{+}$  & 0  &  2.56             \\
12&$^{282}$Ds \cite{hofmann2016} &  8.96   &    1.83$^{+1.39}_{-0.21}$        &  0$^{+}$  & 0$^{+}$  & 0  &  1.38          \\
13&$^{278}$Hs \cite{hofmann2016} &  8.8    &    2.84$^{+1.89}_{-0.21}$     &  0$^{+}$  & 0$^{+}$  & 0  &  1.26             \\
\hline
\multicolumn{8}{c}{$^{299}$120$\rightarrow$$^{295}$Og$\rightarrow$$^{291}$Lv$\rightarrow$$^{287}$Fl$\rightarrow$$^{283}$Cn$\rightarrow$$^{279}$Ds$\rightarrow$$^{275}$Hs$\rightarrow$$^{271}$Sg$\rightarrow$$^{267}$Rf}\\
\hline
14&$^{299}$120 \cite{hofmann2016}  &  13.318  &   -                        &  1/2$^{+}$  & 1/2$^{+}$  & 0  & -6.40   \\
15&$^{295}$Og  \cite{hofmann2016}  &  11.976  &-0.74$^{+0.10}_{-0.22}$    &  1/2$^{+}$  & 1/2$^{+}$  & 0  & -4.10    \\
16&$^{291}$Lv  \cite{hofmann2016}  &  10.847  &-1.89$^{+0.47}_{-0.22}$    &  1/2$^{+}$  & 3/2$^{+}$  & 1  & -1.51      \\
17&$^{287}$Fl  \cite{hofmann2016}  &  10.167  &-0.27$^{+0.07}_{-0.08}$    &  3/2$^{+}$  & 3/2$^{+}$  & 0  & -0.70      \\
18&$^{283}$Cn  \cite{hofmann2016}  &  9.658   &0.65$^{+0.47}_{-0.07}$   &  3/2$^{+}$  & 15/2$^{-}$ & 7  & 2.43         \\
19&$^{279}$Ds  \cite{hofmann2016}  &  9.847   &-0.54$^{+0.12}_{-0.07}$    &  15/2$^{-}$ & 3/2$^{+}$  & 7  & 1.31       \\
20&$^{275}$Hs  \cite{hofmann2016}  &  9.45    &-0.70$^{+1.10}_{-0.15}$    &  3/2$^{+}$  & 3/2$^{+}$  & 0  & -0.43    \\
21&$^{271}$Sg  \cite{hofmann2016}  &  8.66    &1.98$^{+2.28}_{-0.15}$   &  3/2$^{+}$  & 13/2$^{-}$ & 5  & 3.03       \\
22&$^{267}$Rf  \cite{hofmann2016}  &  7.9     &3.66$^{+2.33}_{-0.18}$   &  13/2$^{-}$ & 3/2$^{+}$  & 5  & 4.98         \\
\hline
\hline
\end{tabular}}
\label{decay-chain}
\end{table*}

We have also worked separately on the two regions (i) $N\leq126$ (ii) $N>126$ to apply the half-life formula on the recent (probable) measurements of $\alpha$-decay chains on either side. One of the decay chains is towards the extremely left territory of $\alpha$-decay i.e. the self-conjugate decay chain of $^{108}$Xe referred to as a superallowed $\alpha$-decay to doubly magic $^{100}$Sn ($^{108}$Xe$\rightarrow$$^{104}$Te$\rightarrow$$^{100}$Sn). This decay chain is recently observed \cite{auranen2018} in the ATLAS facility of Argonne National Laboratory and acquires a fair testing ground for both of the proposed formulas. In Table \ref{decay-chain}, we test the accuracy of the proposed formula on this decay chain and also estimate the half-life of the next probable self-conjugate nucleus of this decay chain i.e $^{112}$Ba which is a heavier $N=Z$ $\alpha$-emitter but with possible observation \cite{auranen2018}. Our calculations for this decay chain are found in reasonable agreement endorsing the applicability of proposed formulas in the northwest part of the $\alpha$-decay region. In addition, in the mid-region of $\alpha$-decay mode, most recently, a favored $\alpha$-decay is observed from a new proton-unbound isotope $^{204}$Ac in the China Accelerator Facility for superheavy Elements (CAFE2) \cite{huang2022}. We have tested our formula for this recently observed decay chain along with the estimation of probable decay of $^{208}$Pa: one $\alpha$-particle away from $^{204}$Ac, in Table \ref{decay-chain}. Our results are found in remarkable agreement with the uncertainties and once again demonstrate the utility of the proposed formulas.\par

\begin{figure}[!htbp]
\centering
\includegraphics[width=0.5\textwidth]{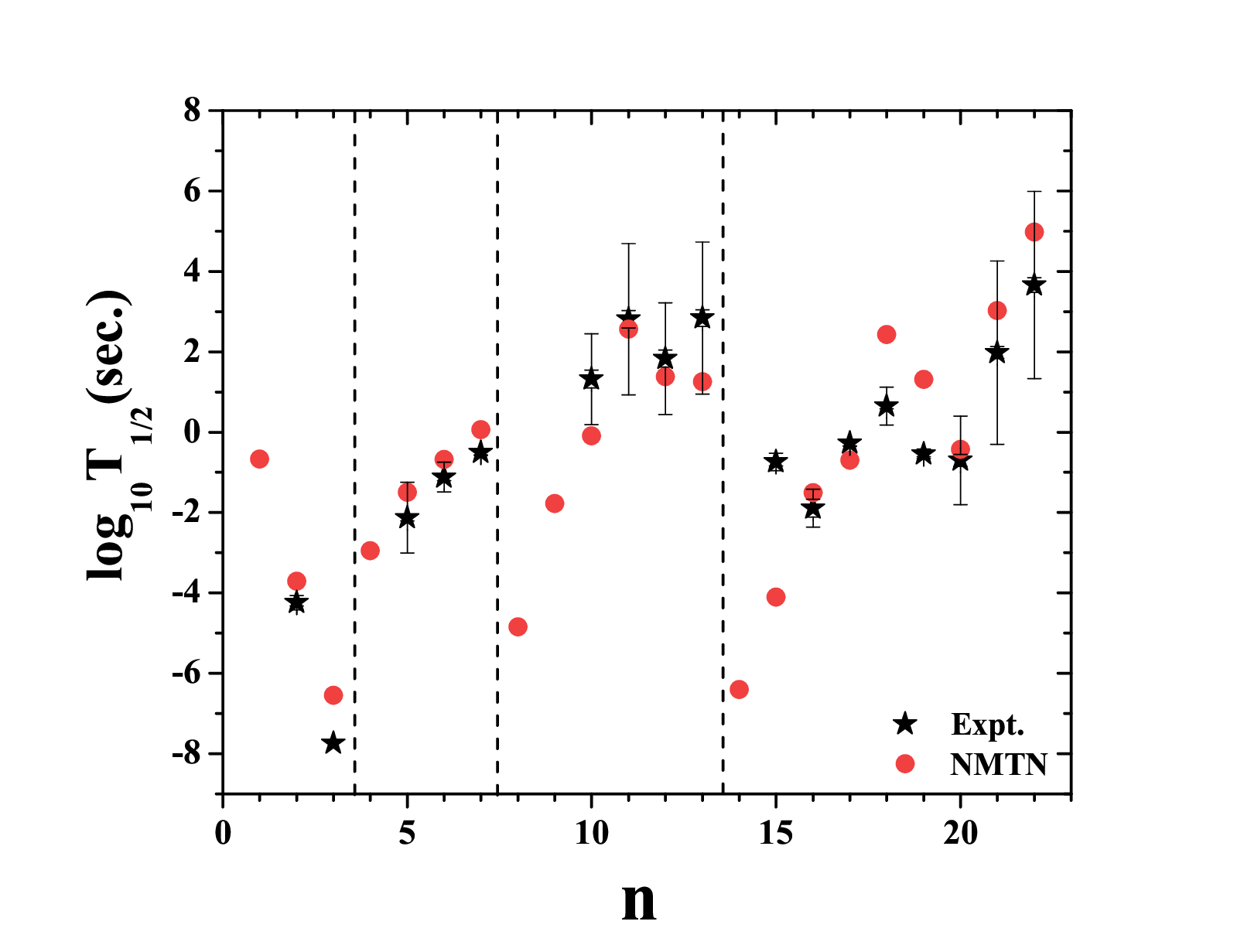} \caption{Logarithm of $\alpha$-decay half-lives calculated by using NMTN formula compared with the experimental half-lives for the decay chains considered in Table \ref{decay-chain}. For the x-axis, n refers to the first column of Table  \ref{decay-chain}.}
\label{fig.comp}
\end{figure}

Other decay chains can be chosen from the north-east part i.e. from the superheavy region to check the pertinence of the provided formula. In view of this, we have chosen tentative $\alpha$-decay chains of $^{298}$Og along with the decay chain of $^{299}$120, where the data are tentatively assigned in Ref. \cite{hofmann2016}. Our estimation of half-lives for these decay chains are shown in Table \ref{decay-chain}, and found to be in reasonable agreement. To depict this agreement, the data of Table \ref{decay-chain} are labeled by an index (n) (please see the first column) and plotted in Fig. \ref{fig.comp}. The majority of the predicted half-lives for the considered decay chains are found within the error bars of the experimental half-lives as can be seen in the different parts of Fig. \ref{fig.comp} for each considered decay chain of the Table \ref{decay-chain}. The deviations in our prediction for a few cases from the assigned tentative data can be attributed to the uncertainties and lesser degree of precision. The estimation of half-lives for $^{112}$Ba, $^{294}$Lv, $^{298}$Og, and $^{299}$120 could be useful for the planning of future experiments. For these nuclei, the Q-values are taken from the WS4 ma there should always be a split up between the training and test data to avoid the chance of overfitting.ss model \cite{ws42014} and mentioned by $\dag$ superscript in the table. For completeness, we have also estimated half-lives of all the nuclei by using the NTMN formula within the range of proton number 111$\leq$Z$\leq$120 and neutron number 160$\leq$N$\leq$184 which are listed in the supplementary material ($supplimentry-material-2-111-Z-120-data.dat$).\par
The present work utilizes a systematic manual approach for the formula selection and its dependencies, which can further be extended by automating the formula creation and selection process using AI-assisted methods such as symbolic regression and others \cite{cornelio2023}.

\section{Conclusion}
Half-lives and preformation factors related to $\alpha$-clusters decay are embedded with microscopic structural information and cast in the form of quite precise empirical formulas which are tested over several other similar and latest formulas using the k-cross validation approach. Shell effects at neutron number N$=$126 are found crucial and used to split the $\alpha$-decay region into two parts. Nevertheless, the role of asymmetry of the parent nucleus and odd-nucleon blocking is found very significant for the accurate determination of half-lives as well as the preformation factor. Because of the applied k-cross validation approach, the reliability of both the formulas is found as unquestioned, and hence, half-lives for decay-chains of $^{112}$Ba and $^{208}$Pa complying with recently observed chains, together with decay chains of the most anticipated superheavy nuclei $^{298}$Og and $^{299}$120, are estimated with a reasonable agreement. To utilize these formulas further, values of calculated half-lives for (i) dataset of 349 nuclei and (ii) nuclei with 111$\leq$Z$\leq$120 and neutron number 160$\leq$N$\leq$184, are provided as supplementary materials.
As a takeaway message to the readers, it is pointed out here that the merit of any formula should be compared with others, only if all the formulas are fitted on the same dataset, and fundamentally

\section*{Acknowledgments}
GS acknowledges the support provided by SERB (DST), Govt. of India under SIR/2022/000566, and would like to thank Prof. Nils Paar for his kind hospitality at the University of Zagreb, Croatia. Authors are indebted to Riya Sailani, UOR, Jaipur, India for the discussions.

\end{document}